\documentstyle[12pt]{article}
\newcommand{\eq}{\begin{equation}}
\newcommand{\en}{\end{equation}}
\newcommand{\eqn}{\begin{eqnarray}}
\newcommand{\enn}{\end{eqnarray}}
\newcommand{\CR}{\nonumber \\}
\newcommand{\pa}{\partial}

\newcommand{\D}{\delta}          

\newcommand{\E}{\epsilon}
\newcommand{\Lm}{\Lambda}
\newcommand{\p}{\Phi}

\newcommand{\lm}{\lambda}
\newcommand{\s}{\sigma}

\begin{document}

\begin{titlepage}
\null
\begin{flushright} 
hep-th/9701009  \\
UTHEP-356 \\
December, 1996
\end{flushright}
\vspace{0.5cm} 
\begin{center}
{\Large 
N=2 Curves and a Coulomb Phase in N=1 SUSY Gauge Theories

with Adjoint and Fundamental Matters
\par}
\lineskip .75em
\vskip2.5cm
\normalsize
{\large Takuhiro Kitao${}^1$, Seiji Terashima${}^2$ and Sung-Kil Yang${}^2$} 
\vskip 1.5em
${}^1${\it Institute of Physics, University of Tokyo, 
Komaba, Meguro-ku, Tokyo 153, Japan}
\vskip 1.2em
${}^2${\it Institute of Physics,  University of Tsukuba,  Ibaraki 305,  Japan}
\vskip3cm
{\bf Abstract}
\end{center} \par
We study low-energy 
effective superpotentials for the phase with a confined photon
in $N=1$ supersymmetric gauge theories with an adjoint matter $\Phi$ and 
fundamental flavors $Q, \tilde Q$.
Arbitrary classical gauge groups are considered. 
The results are used to derive the hyperelliptic curves which describe 
the Coulomb phase of $N=2$ supersymmetric QCD with classical gauge groups. 
These curves are in agreement with those proposed earlier by several
authors. Our results also produce the curves relevant to describe
the Coulomb phase of $N=1$ theories with a superpotential of the form 
$\tilde{Q} \Phi^{l} Q$.
\end{titlepage}

\baselineskip=0.7cm


In the last three years, there has been a significant development
in understanding four-dimensional supersymmetric gauge theories 
\cite{Se},\cite{SeWi},\cite{Se2}.
In particular new light is shed on mechanism for confinement in view of
condensation of massless solitons in the vacua of the Coulomb phase 
in $N=2$ supersymmetric gauge theories. It is important that
these massless solitons appear at the singularities of moduli space 
of vacua and the superpotential which breaks $N=2$ supersymmetry to $N=1$ 
causes condensation. Conversely, from the analysis of the
confining phase of $N=1$ theories with a suitable superpotential,
we can identify the singular points in the Coulomb phase 
of $N=2$ theories. Thus it becomes possible to
construct the $N=2$ Seiberg-Witten curves 
since the $N=2$ curves are determined almost completely 
by the singularity conditions. This idea has been successfully applied to 
$N=2$ supersymmetric Yang-Mills theories with the classical groups as well as
$G_2$, and $N=2$ supersymmetric $SU(N_c)$ QCD \cite{InSe}-\cite{Ki}.

In this paper, we extend our analysis to describe the Coulomb phase of 
$N=1$ supersymmetric gauge theories with an adjoint matter field $\Phi$ and
fundamental flavors $Q, \tilde Q$. A tree-level superpotential consists of 
the Yukawa-like term $\tilde{Q} \Phi^{l} Q$ in addition to the Casimir
terms built out of $\Phi$, and we shall consider arbitrary classical gauge
groups. In the appropriate limit the theory is reduced to $N=2$ supersymmetric 
QCD. We derive the low-energy effective superpotential for the phase with a
single confined photon and obtain
the hyperelliptic curves which describe the Coulomb phase of the theory.
In the $N=2$ limit we will show that these curves agree with the results of 
\cite{HO}-\cite{ArSh}, and hence our results are also viewed as a non-trivial
test of the $N=2$ curves proposed previously.


We start with discussing $N=1$ $SU(N_c)$ supersymmetric gauge theory
with an adjoint matter field $\p$, $N_f$ flavors of fundamentals $Q$ and
anti-fundamentals $\tilde{Q}$ to explain our method in this paper.
We take a tree-level superpotential 
\eq
W=\sum_{n=1}^{N_c} g_n u_n
+\sum_{l=0}^{r} {\rm Tr}_{N_f} \, \lm_l \, \tilde{Q} \p^l Q, 
\hskip10mm  u_n=\frac{1}{n} {\rm Tr}\,  \p^n ,
\label{r1}
\en
where ${\rm Tr}_{N_f} \, \lm_l \, \tilde{Q} \p^l Q = 
\sum_{i,j=1}^{N_f} (\lm_l)^i_j  \tilde{Q}_i \p^l Q^j$ and $r \leq N_c-1$.
If we set $(\lm_0)^i_j=m^i_j$ with $[m, m^{\dagger}]=0$, 
$(\lm_1)^i_j=\D^i_j , \, 
(\lm_l)^i_j=0 $ for $l>1$ and all $g_i=0$, eq.(\ref{r1}) recovers the
superpotential in $N=2$ $SU(N_c)$ supersymmetric QCD with quark mass $m$.
The second term in (\ref{r1}) was considered in a recent work \cite{Ka}.

Let us focus on the classical vacua with $ Q=\tilde{Q}=0$ and an
unbroken $SU(2) \times U(1)^{N_c-2}$ symmetry
which means $\p  ={\rm diag} (a_1, a_1, a_2, a_3, \cdots , a_{N_c-1})$ 
up to gauge transformations. 
(Note that the superpotential (\ref{r1}) has no classical vacua with unbroken 
$U(1)^{N_c-1}$.) In this vacuum, we will evaluate semiclassicaly 
the low-energy effective superpotential.
Our procedure is slightly different from that adopted in \cite{ElFoGiInRa}
upon treating $Q$ and $\tilde{Q}$.
We investigate the tree-level parameter region where
the Higgs mechanism occurs at very high energies and 
the adjoint matter field $\p$ is quite heavy.
Then the massive particles are integrated out and 
the scale matching relation becomes
\eq
{\Lm_L}^{6-N_f} = g_{N_c}^2 \Lm^{2 N_c-N_f},
\label{sumatch}
\en
where $\Lm$ is the
dynamical scale of high-energy $SU(N_c)$ theory with $N_f$ flavors
and $\Lm_L$ is the scale of low-energy $SU(2)$ theory with $N_f$ flavors.
Eq.(\ref{sumatch}) is derived by following the $SU(N_c)$ 
Yang-Mills case \cite{ElFoGiInRa} while taking into account the existence of 
fundamental flavors at low energies \cite{KSS}.

The semiclassical superpotential in low-energy $SU(2)$ theory with $N_f$ 
flavors reads
\eq
W=\sum_{n=1}^{N_c} g_n u_n^{cl} +
\sum_{l=0}^{r} a_1^l \, {\rm Tr}_{N_f} \, \lm_l \, \tilde{Q} Q
\label{w1}
\en
which is obtained by substituting the classical values of $\p$
and integrating out all the fields except for those coupled to 
the $SU(2)$ gauge boson.
Here, the constraint ${\rm Tr} \p^{cl}=a_1+\sum_{i=1}^{N_c-1} a_i=0$ and
the classical equation of motion $\sum_{i=1}^{N_c-1} a_i
=-g_{N_c-1}/g_{N_c}$ yield \cite{Ki}
\eq
a_1= \frac{g_{N_c-1}}{g_{N_c}}.
\en
Below the flavor masses which can be read off from the superpotential 
(\ref{w1}), the low-energy theory becomes $N=1$ $SU(2)$ Yang-Mills theory 
with the superpotential
\eq
W=\sum_{n=1}^{N_c} g_n u_n^{cl}.
\label{w2}
\en
This low-energy theory 
has the dynamical scale $\Lm_{YM}$ which is related to $\Lm$ through
\eq
{\Lm_{YM}}^{6} = {\rm det} \left ( \sum_{l=0}^{r} \lm_l a_1^l \right ) \,
g_{N_c}^2 \Lm^{2 N_c-N_f}.
\label{sc1}
\en

As in the previous literatures \cite{ElFoGiInRa},\cite{TeYa}
we simply assume here that 
the superpotential (\ref{w2}) and the scale matching relation (\ref{sc1}) 
are exact for any values of the tree-level parameters.
Now we add to (\ref{w2}) a dynamically generated piece which arises from 
gaugino condensation in $SU(2)$ Yang-Mills theory.
The resulting effective superpotential $W_L$ where all the matter fields have
been integrated out is thus given by
\eqn
W_L & = & \sum_{n=1}^{N_c} g_n u_n^{cl} \pm 2 \Lm_{YM}^3 \CR
  & = & \sum_{n=1}^{N_c} g_n u_n^{cl} \pm 2 g_{N_c} \sqrt{A(a_1)}
\label{w3}
\enn
with $A$ being defined as $A(x) \equiv \Lm^{2 N_c-N_f} \,
{\rm det} \left ( \sum_{l=0}^{r} \lm_l x^l \right ) $. From 
$\langle u_n \rangle = \partial W_L/\partial g_n$ we find
\eq
\langle u_n \rangle = u_n^{cl} (g) \pm \D_{n,N_c-1} 
\frac{A'(a_1)}{\sqrt{A(a_1)}}
\pm \D_{n,N_c}  \frac{1}{\sqrt{A(a_1)}} \left ( 2 A(a_1) -a_1 A'(a_1) \right).
\label{v1}
\en

If we define a hyperelliptic curve
\eq
y^2= P(x)^2 -4 A(x),
\label{c1}
\en
where $P(x)=\left \langle {\rm det} \left ( x- \p \right ) \right \rangle$
is the characteristic equation of $\p$, the curve is quadratically 
degenerate at the vacuum expectation values (\ref{v1}).
This can be seen by plugging (\ref{v1}) in $P(x)$
\eq
P(x)=P_{cl} (x) \mp x  \frac{A'(a_1)}{\sqrt{A(a_1)}}
\mp \frac{1}{\sqrt{A(a_1)}} \left ( 2 A(a_1) -a_1 A'(a_1) \right),
\en
where $P_{cl}(x)={\rm det} \left ( x- \p_{cl} \right )$, and hence
\eq
P(a_1)= \mp 2 {\sqrt{A(a_1)}} \, , \hskip10mm
P'(a_1)= \mp \frac{A'(a_1)}{\sqrt{A(a_1)}}.
\en
Then the degeneracy of the curve is confirmed by checking
$ y^2|_{x=a_1}=0$ and 
$ \frac{\pa}{\pa x} y^2 |_{x=a_1} = 0$.

The transition points from the confining to the Coulomb phase are reached by
taking the limit $g_{i} \rightarrow 0$ while keeping the ratio $g_i/g_{j}$ 
fixed \cite{ElFoGiInRa}. 
These points correspond to the singularities in the moduli space.
Therefore the curve (\ref{c1}) is regarded as 
the curve relevant to describe the Coulomb phase of the theory with the
tree-level superpotential 
$W=\sum_{l=0}^{r} {\rm Tr}_{N_f} \, \lm_l \, \tilde{Q} \p^l Q$.
Indeed, the curve (\ref{c1}) agrees with the one obtained in \cite{Ka}.
Especially in the parameter region that has $N=2$ supersymmetry,
we find agreement with the curves for $N=2$ $SU(N_c)$ QCD with 
$N_f<2 N_c-1$ \cite{HO},\cite{ArPlSh},\cite{ArSh}.\footnote[2]{For 
$N_f=2 N_c-1$ an instanton may generate a mass term and shift
the bare quark mass in $A(x)$. If we include this effect the curve (\ref{c1})
completely agrees with the result in \cite{ArSh}.}


The procedure discussed above can be also applied to the other classical
gauge groups.
Let us consider $N=1$ $SO(2 N_c)$ supersymmetric gauge theory
with an adjoint matter field $\p$ 
which is an antisymmetric $2 N_c \times 2 N_c$ tensor, and $2 N_f$ flavors 
of fundamentals $Q$. We assume a tree-level superpotential 
\eq
W=\sum_{n=1}^{N_c-2} g_{2 n} u_{2 n} + g_{2 (N_c-1)} s_{N_c-1}+\lm v
+{1\over 2}\sum_{l=0}^{r} {\rm Tr}_{2 N_f} \, \lm_l \, Q \p^l Q, 
\label{w4}
\en
where $r \leq 2 N_c-1$,
\eq
u_{2 n} =\frac{1}{2 n} {\rm Tr}\, \Phi^{2 n}, \hskip10mm
v ={\rm Pf}\, \Phi=\frac{1}{2^{N_c} N_c !} \E_{i_1 i_2 j_1 j_2 \cdots}
\Phi^{i_1 i_2} \Phi^{j_1 j_2} \cdots  
\en
and
\eq
ks_k+\sum_{i=1}^k i s_{k-i} u_{2i}=0, \hskip10mm s_0=-1, 
\hskip10mm k=1,2,\cdots .
\en
Here, ${}^t \lm_l=(-1)^l \lm_l$ and the $N=2$ supersymmetry is present
when we set $(\lm_0)^i_j=m^i_j$ where $[m, m^{\dagger}]=0$, 
$(\lm_1)^i_j={\rm diag} (i \s_2,i \s_2, \cdots )$ with 
$\s_{2} = \pmatrix{0 & -i \cr i & 0}, \, 
(\lm_l)^i_j=0 $ for $l>1$ and all $g_i=0$. 

As in the case of $SU(N_c)$, 
we concentrate on the unbroken $SU(2) \times U(1)^{N_c-1}$ vacua with
$\p  ={\rm diag} 
(a_1 \s_2, a_1 \s_2 , a_2 \s_2, a_3 \s_2, \cdots , a_{N_c-1} \s_2)$
and $Q=0$.
By virtue of using $s_{N_c}$ instead of $u_{2 N_c}$ in (\ref{w4})
the degenerate eigenvalue of $\p_{cl}$ is expressed by $g_i$
\eq
a_1^2=\frac{g_{2(N_c-2)}}{g_{2(N_c-1)}}
\en
as found for the $SO(2 N_c+1)$ case \cite{TeYa}.
Note that the superpotential (\ref{w4}) has
no classical vacua with unbroken $SO(4) \times U(1)^{N_c-1}$ 
when $g_{2 (N_c-2)} \neq 0$.
We also note that 
the fundamental representation of $SO(2 N_c)$ is decomposed into 
two fundamental representations of $SU(2)$ under the above embedding.
It is then observed that 
the scale matching relation between the high-energy $SO(2 N_c)$ scale $\Lm$ and
the scale $\Lm_L$ of low-energy $SU(2)$ theory with $2 N_f$ fundamental 
flavors is given by
\eq
{\Lm_L}^{6-2 N_f} = g_{2(N_c-1)}^2 \Lm^{4( N_c-1) -2 N_f}.
\en

The superpotential for low-energy $N=1$ $SU(2)$ QCD with $2 N_f$ flavors
can be obtained in a similar way to the $SU(N_c)$ case, but the duplication
of the fundamental flavors are taken into consideration. 
After some manipulations it turns out that the superpotential for 
low-energy $N=1$ $SU(2)$ QCD with $2 N_f$ flavors is written as
\eq
W=\sum_{n=1}^{N_c-2} g_{2 n} u_{2 n}^{cl} 
+ g_{2 (N_c-1)} s_{N_c-1}^{cl}+\lm v^{cl}
+\sum_{l=0}^{r} a_1^l {\rm Tr}_{2 N_f} \, \lm_l \, \widetilde{{\bf Q}} {\bf Q},
\label{w5}
\en
where
\eq
{\bf Q}^j = {1\over \sqrt{2}} \pmatrix{Q^j_1-iQ^j_2 \cr
                     Q^j_3-iQ^j_4},     \hskip10mm
\widetilde{\bf Q}_j= {1\over \sqrt{2}}\pmatrix{Q^j_1+iQ^j_2 \cr
                          Q^j_3+iQ^j_4}.
\en
Upon integrating out the $SU(2)$ flavors we have
the scale matching  between 
$\Lm$ and $\Lm_{YM}$ for $N=1$ $SU(2)$ Yang-Mills theory 
\eq
{\Lm_{YM}}^{6} = {\rm det} \left ( \sum_{l=0}^{r} \lm_l a_1^l \right ) \,
g_{2(N_c-1)}^2 \Lm^{4( N_c-1)-2 N_f},
\label{sc2}
\en
and we get the effective superpotential 
\eqn
W_L & = & \sum_{n=1}^{N_c-2} g_n u_n^{cl} 
+ g_{2 (N_c-1)} s_{N_c-1}^{cl}+\lm v^{cl}
\pm 2 \Lm_{YM}^3 \CR
  & = & \sum_{n=1}^{N_c-2} g_n u_n^{cl} 
+ g_{2 (N_c-1)} s_{N_c-1}^{cl}+\lm v^{cl} \pm 2 g_{2(N_c-1)} \sqrt{A(a_1)},
\label{wso}
\enn
where $A$ is defined by $A(x) \equiv \Lm^{4( N_c-1)-2 N_f} \,
{\rm det} \left ( \sum_{l=0}^{r} \lm_l x^l \right ) =A(-x)$.

The vacuum expectation values of gauge invariants are obtained from $W_L$ as
\eqn
\langle s_{ n} \rangle & =&  s_{ n}^{cl} (g) 
\pm \D_{n,N_c-2} \frac{A'(a_1)}{\sqrt{A(a_1)}}
\pm \D_{n,N_c-1}  \frac{1}{\sqrt{A(a_1)}} 
\left ( 2 A(a_1) -a_1^2 A'(a_1) \right),    \CR
\langle v \rangle & =&  v^{cl} (g),
\label{vso}
\enn
where $A'(x)=\frac{\pa}{\pa x^2} A(x)$.
It is now easy to see that a curve 
\eq
y^2=P(x)^2-4 x^4 A(x) 
\en
with $P(x)=\left \langle {\rm det} \left ( x-\p \right ) \right \rangle$
is degenerate at these values of $\langle s_n \rangle, \, \langle v \rangle$, 
and reproduces the known $N=2$ curve \cite{H}, \cite{ArSh}.


The only difference between $SO(2N_c)$ and $SO(2 N_c+1)$ is that 
the gauge invariant ${\rm Pf}\, \p$ vanishes for $SO(2 N_c+1)$.
Thus, taking a tree-level superpotential 
\eq
W=\sum_{n=1}^{N_c-1} g_{2 n} u_{2 n} + g_{2 N_c} s_{N_c}
+{1\over 2} \sum_{l=0}^{r} {\rm Tr}_{2 N_f} \, \lm_l \, Q \p^l Q , 
\hskip10mm  r \leq 2N_c ,
\label{wsoo}
\en
we focus on the unbroken $SU(2) \times U(1)^{N_c-1}$ vacuum which has 
the classical expectation values 
$\p  ={\rm diag} (a_1 \s_2, a_1 \s_2 , a_2 \s_2, \cdots , a_{N_c-1} \s_2,0)$ 
and $Q=0$ \cite{TeYa}.
As in the $SO(2 N_c)$ case we make use of
the scale matching relation between the high-energy scale $\Lm$ and
the low-energy $N=1$ $SU(2)$ Yang-Mills scale $\Lm_{YM}$ 
\eq
{\Lm_{YM}}^{6} = {\rm det} \left ( \sum_{l=0}^{r} \lm_l a_1^l \right ) \,
g_{2 N_c} g_{2(N_c-1)} \Lm^{2(2 N_c-1- N_f)}.
\label{sc3}
\en
As a result we find the effective superpotential 
\eqn
W_L & = & \sum_{n=1}^{N_c-1} g_{2n} u_n^{cl} + g_{2 N_c} s_{N_c}^{cl}
\pm 2 \Lm_{YM}^3 \CR
  & = & \sum_{n=1}^{N_c-1} g_{2n} u_n^{cl} + g_{2 N_c} s_{N_c}^{cl}
\pm 2 \sqrt{g_{2 N_c} g_{2(N_c-1)} A(a_1)},
\label{wsoo2}
\enn
where $A$ is defined as $A(x) \equiv \Lm^{2( 2 N_c-1- N_f)}\,
{\rm det} \left ( \sum_{l=0}^{r} \lm_l x^l \right )$.

Noting the relation $a_1^2=g_{2(N_c-1)}/g_{2 N_c}$ \cite{TeYa}
we calculate the vacuum expectation values of gauge invariants 
\eqn
\langle s_{ n} \rangle =  s_{ n}^{cl} (g) 
& \pm & \D_{n,N_c-1} \frac{1}{\sqrt{A(a_1)}} 
\left (  \frac{A(a_1)}{a_1}+a_1 A'(a_1) \right) \CR
& \pm & \D_{n,N_c}  \frac{1}{\sqrt{A(a_1)}} 
\left ( a_1 A(a_1) -a_1^3 A'(a_1) \right).
\label{vsoo}
\enn
For these $\langle s_n \rangle$ we observe 
the quadratic degeneracy of the curve
\eq
y^2=\left( \frac{1}{x} P(x) \right )^2-4 x^2 A(x) ,
\en
where $P(x)=\left \langle {\rm det} \left ( x-\p \right ) \right \rangle$.
In the $N=2$ limit we see agreement with the curve constructed in 
\cite{H},\cite{ArSh}. The confining phase superpotential for the $SO(5)$ gauge
group was obtained also in \cite{LaPiGi}.


Let us now turn to $Sp(2 N_c)$ gauge theory.
We take for matter content an adjoint field $\p$
and $2 N_f$ fundamental fields $Q$.
The $2N_c \times 2N_c$ tensor $\p$ is subject to ${}^t \Phi = J \Phi J $ with
$J={\rm diag}(i\sigma_2, \cdots, i\sigma_2)$.
Our tree-level superpotential reads
\eq
W=\sum_{n=1}^{N_c-1} g_{2 n} u_{2 n} + g_{2 N_c} s_{N_c}
+{1\over 2} \sum_{l=0}^{r} {\rm Tr}_{2 N_f} \, \lm_l \, Q J \p^l Q,
\label{wsp}
\en
where ${}^t \lm_l=(-1)^{l+1} \lm_l$ and $r \leq 2N_c-1$.
The classical vacuum with the unbroken $SU(2) \times U(1)^{N_c-1}$ gauge group
corresponds to
\eq
J \Phi=
{\rm diag}(\s_{1}a_1,\; \s_{1}a_1,\; \s_{1}a_2,\cdots,\s_{1}a_{N_c-1}) ,
\hskip10mm \s_{1}
=\left( \begin{array}{cc} 0 &  1 \\ 1 & 0 \end{array} \right ),
\en
where $a_1^2=g_{2(N_c-1)}/g_{2N_c}$.
The scale $\Lm_L$ for low-energy $SU(2)$ theory with $2 N_f$ flavors
is expressed as \cite{TeYa}
\eq
{\Lm_L}^{6-2 N_f} = \left( \frac{g_{2 N_c}^2}{g_{2 (N_c-1)}} \right)^2 
\Lm^{2(2N_c+2- N_f)}.
\label{embed}
\en

There exists a subtle point in the analysis of $Sp(2 N_c)$ theory. When
$Sp(2 N_c)$ is broken to $SU(2) \times U(1)^{N_c-1}$ the instantons in
the broken part of the gauge group play a role since the index of the embedding
of the unbroken $SU(2)$ in $Sp(2 N_c)$ is larger than one 
(see eq.(\ref{embed})) \cite{InSe2},\cite{Aff}. 
The possible instanton contribution to $W_L$ will be of the 
same order in $\Lm$ as low-energy $SU(2)$ gaugino condensation.
Therefore even in the lowest quantum corrections
the instanton term must be added to $W_L$.

For clarity we begin with discussing $Sp(4)$ Yang-Mills theory.
In this theory by the symmetry and holomorphy
the effective superpotential is determined to take the form 
$W_L=f \left( \frac{g_4}{g_2} \Lm^2 \right) \frac{g_4^2}{g_2} \Lm^6$
with $f$ being certain holomorphic function.
If we assume that there is  only one-instanton effect,
the precise form of $W_L$ including the low-energy gaugino 
condensation effect may be given by
\eq
W_L= 2 \frac{g_4^2}{g_2} \Lm^6 \pm 2 \frac{g_4^2}{g_2} \Lm^6,
\en 
as in the case of $SO(4) \simeq SU(2) \times SU(2)$ breaking to 
the diagonal $SU(2)$. This is due to the fact $Sp(4) \simeq SO(5)$ and the
natural embedding of $SO(4)$ in $SO(5)$. Our low-energy $SU(2)$ gauge
group is identified with the one diagonally embedded in 
$SO(4) \simeq SU(2) \times SU(2)$ \cite{InSe2},\cite{ILS}. Accordingly, 
in $Sp(2 N_c)$ Yang-Mills theory, we first make the matching at the scale of 
$Sp(2 N_c)/Sp(4)$ $W$ bosons by taking all the $a_1-a_i$ large. 
Then the low-energy superpotential is found to be
\eq
W_L=W_{cl}+2  \frac{g_{2 N_c}}{a_1^2} \Lm^{2(N_c+1)}
\pm 2 \frac{g_{2 N_c}}{a_1^2} \Lm^{2(N_c+1)}.
\en 

Let us turn on the coupling to fundamental flavors $Q$ and evaluate
the instanton contribution. When flavor masses vanish there is a global
$O(2N_f) \simeq SO(2N_f)\times {\bf Z}_2$ symmetry. The couplings $\lambda_l$
and instantons break a ``parity'' symmetry ${\bf Z}_2$. We treat this 
${\bf Z}_2$ as unbroken by assigning odd parity to the instanton factor
$\Lm^{2 N_c+2- N_f}$ and $O(2N_f)$ charges to $\lambda_l$. Symmetry
consideration then leads to the one-instanton factor proportional to
$B(a_1)$ where
\eq
B(x)=\Lm^{2 N_c+2- N_f} {\rm Pf}
\left(\sum_{l\, {\rm even}} \lambda_{l}x^{l}\right).
\en
Note that $B(x)$ is parity invariant since Pfaffian has odd parity.
Thus the superpotential for low-energy $N=1$ $SU(2)$ QCD with $2 N_f$ flavors
including the instanton effect turns out to be
\eq
W=\sum_{n=1}^{N_c-1} g_{2 n} u_{2 n}^{cl} + g_{2 N_c} s_{N_c}^{cl}
+\sum_{l=0}^{r} a_1^l {\rm Tr}_{2 N_f} \, \lm_l \, \widetilde{{\bf Q}} {\bf Q}
+2  \frac{g_{2 N_c}^2}{g_{2 (N_c-1)}} B(a_1),
\label{wsp2}
\en
where
\eq
{\bf Q}^j = \pmatrix{Q^j_1 \cr
                     Q^j_3},     \hskip10mm
\widetilde{\bf Q}_j= \pmatrix{Q^j_2 \cr
                          Q^j_4}.
\en
When integrating out the $SU(2)$ flavors,
the scale matching relation between $\Lm$ and the scale $\Lm_{YM}$ 
of $N=1$ $SU(2)$ Yang-Mills theory becomes
\eq
{\Lm_{YM}}^{6} = {\rm det} \left ( \sum_{l=0}^{r} \lm_l a_1^l \right ) \,
\left(  \frac{g_{2 N_c}^2}{g_{2 (N_c-1)}} \right)^2 
\Lm^{2(2N_c+2- N_f)},
\label{sc4}
\en
and we finally obtain the effective superpotential 
\eqn
W_L & = & \sum_{n=1}^{N_c-1} g_n u_n^{cl} + g_{2 N_c} s_{N_c}^{cl}
\pm 2 \Lm_{YM}^3 +2  \frac{g_{2 N_c}^2}{g_{2 (N_c-1)}} B(a_1) \CR
  & = & \sum_{n=1}^{N_c-1} g_n u_n^{cl} + g_{2 N_c} s_{N_c}^{cl}
+2 \frac{g_{2 N_c}^2}{g_{2 (N_c-1)}}  
\left(B(a_1) \pm \sqrt{A(a_1)} \right),
\label{wsp3}
\enn
where $A(x) \equiv \Lm^{2( 2 N_c+2- N_f) }\,
{\rm det} \left ( \sum_{l=0}^{r} \lm_l x^l \right )$.

The gauge invariant expectation values $\langle s_n \rangle$ are 
\eqn
\langle s_{ n} \rangle =  s_{ n}^{cl} (g) 
& +& \D_{n,N_c-1} \frac{1}{a_1^4} \left(
-2 B(a_1)+2a_1^2B'(a_1)  \pm \frac{1}{\sqrt{A(a_1)}} 
\left( -2 A(a_1) + a_1^2 A'(a_1) \right) \right) \CR
& + & \D_{n,N_c}  \frac{1}{a_1^2} \left( 
4 B(a_1)-2a_1^2B'(a_1) \pm  \frac{1}{\sqrt{A(a_1)}} 
\left( 4 A(a_1)  - a_1^2 A'(a_1) \right) \right).
\label{vsp}
\enn
Substituting these into a curve 
\eq
x^2 y^2 = \left( x^2 P(x) +2 B(x) \right)^2-4 A(x),
\label{spcurve}
\en
we see that the curve is degenerate at (\ref{vsp}). In this case too 
our result (\ref{spcurve}) agrees with the $N=2$ curve obtained in \cite{ArSh}.


Using the technique of confining phase superpotential we have determined
the curves describing the Coulomb phase of $N=1$ supersymmetric gauge theories
with adjoint and fundamental matters with classical gauge groups.
In the $N=2$ limit our results recover the curves for the Coulomb phase in
$N=2$ QCD. For the gauge group $Sp(2N_c)$, in particular, we have observed
that taking into account the instanton effect in addition to $SU(2)$ gaugino
condensation is crucial to obtain the effective superpotential for the phase
with a confined photon. This explains in terms of $N=1$ theory a peculiar
feature of the $N=2$ $Sp(2N_c)$ curve when compared to the $SU(N_c)$ and
$SO(N_c)$ cases.


\vskip10mm
We thank K. Ito for helpful discussions on the instanton calculus. 
TK would like to thank T. Hotta for useful discussions.
The work of ST is supported by JSPS Research Fellowship for Young 
Scientists. The work of SKY is supported in part by the
Grant-in-Aid for Scientific Research on Priority Area 213 
``Infinite Analysis'', the Ministry of Education, Science and Culture, Japan.

\newpage


\end{document}